\documentclass[11pt]{article}
\usepackage{moriond,epsfig}
\usepackage{epstopdf}
\usepackage{lineno}

\newcommand{\bpm} {\ensuremath{\mathrm{B^{\pm}}}}
\newcommand{\bd} {\ensuremath{\mathrm{B^{0}}}}

\newcommand{\bsmumu}              {\ensuremath{\mathrm{B^{0}_{s}}\to\mu^{+}\mu^{-}}}
\newcommand{\bs}              {\ensuremath{\mathrm{B^{0}_{s}}}}

\newcommand{\dgs}              {\ensuremath{\mathrm{\Delta\Gamma_{s}}}}

\newcommand{\bdmumu}              {\ensuremath{\mathrm{B^{0}}\to\mu^{+}\mu^{-}}}
\newcommand{\jpsi}              {\ensuremath{\mathrm{\mathrm{J}/\psi}}}
\newcommand{\jpsitwos}              {\ensuremath{\mathrm{\psi(2S)}}}
\newcommand{\us}              {\ensuremath{\mathrm{\Upsilon(1S)}}}
\newcommand{\utwos}              {\ensuremath{\mathrm{\Upsilon(2S)}}}
\newcommand{\uthrees}              {\ensuremath{\mathrm{\Upsilon(3S)}}}

\newcommand{\chic}              {\ensuremath{\mathrm{\chi_{c}}}}
\newcommand{\chib}              {\ensuremath{\mathrm{\chi_{b}}}}
\newcommand{\pt}              {\ensuremath{\mathrm{p_{T}}}}

\newcommand{\bsjpsiphi}               {\ensuremath{\mathrm{B^{0}_{s}}\to \mathrm{J}/\psi \phi}}
\newcommand{\bjpsiks}               {\ensuremath{\mathrm{B^{0}}\to \mathrm{J}/\psi \mathrm{K}^{0}_{\mathrm{s}}}}
\newcommand{\lambb}		{\ensuremath{\mathrm{\Lambda_{b}}}}
\newcommand{\lambdab}               {\ensuremath{\mathrm{\Lambda_{b}  \to \mathrm{J}/\psi \Lambda}}}
\newcommand{\chidecay}               {\ensuremath{\mathrm{\chi_{b}\to \Upsilon(ks) \, \gamma}}}

\def\be{\begin{equation}}
\def\ee{\end{equation}}
\def\bea{\begin{eqnarray}}
\def\eea{\end{eqnarray}}

\begin{document}
\vspace*{4cm}
\title{Heavy Flavor Measurements in ATLAS and CMS}
\author{ J.~Schieck on behalf of the ATLAS and CMS Collaboration}
\address{Ludwig-Maximilians-Universit\"at
  M\"unchen , Am Coulombwall 1, 85748 Garching, Germany  and\\
Excellence Cluster Universe, Boltzmannstrasse 2, 85748 Garching, Germany}
\maketitle\abstracts{
We present heavy flavor measurements performed by the ATLAS and CMS Collaborations
with data collected at the Large Hadron Collider. The production mechanism of heavy flavor
hadrons is discussed as well as lifetime measurements and searches for the rare 
decay \bsmumu. The large available statistics of about 5 fb$^{-1}$ per experiment collected during the year 2011 together 
with the excellent detector performance allows to perform competitive heavy flavor measurements.}
\section{Introduction}
The production and the decay of hadrons containing a b-quark provide an excellent laboratory 
to study the strong as well as the weak interaction of the Standard Model of
Particle Physics. The large b-quark production cross-section in proton-proton collisions 
at the Large Hadron Collider (LHC) offers the possibility of performing heavy flavor measurements 
with an unprecedented statistical accuracy. Here we present heavy flavor results obtained with data taken by the
ATLAS~\cite{Aad:2008zzm} and CMS~\cite{Chatrchyan:2008aa} detectors in the year 2010 and 2011. \par
In section~\ref{ExpSetup} we will describe the experimental setup together with the collected data 
used by the measurements, in section~\ref{BHadProd} B-hadron production measurements are summarized,
in section~\ref{Lifetime} lifetime measurements are discussed and finally in section~\ref{RareDecays} the search
for the rare B-hadron decay \bsmumu\ is presented. We conclude with a summary in section~\ref{summary}.
\section{Experimental Setup}
\label{ExpSetup}
\begin{figure}[ht]
\includegraphics[width=0.5\textwidth]{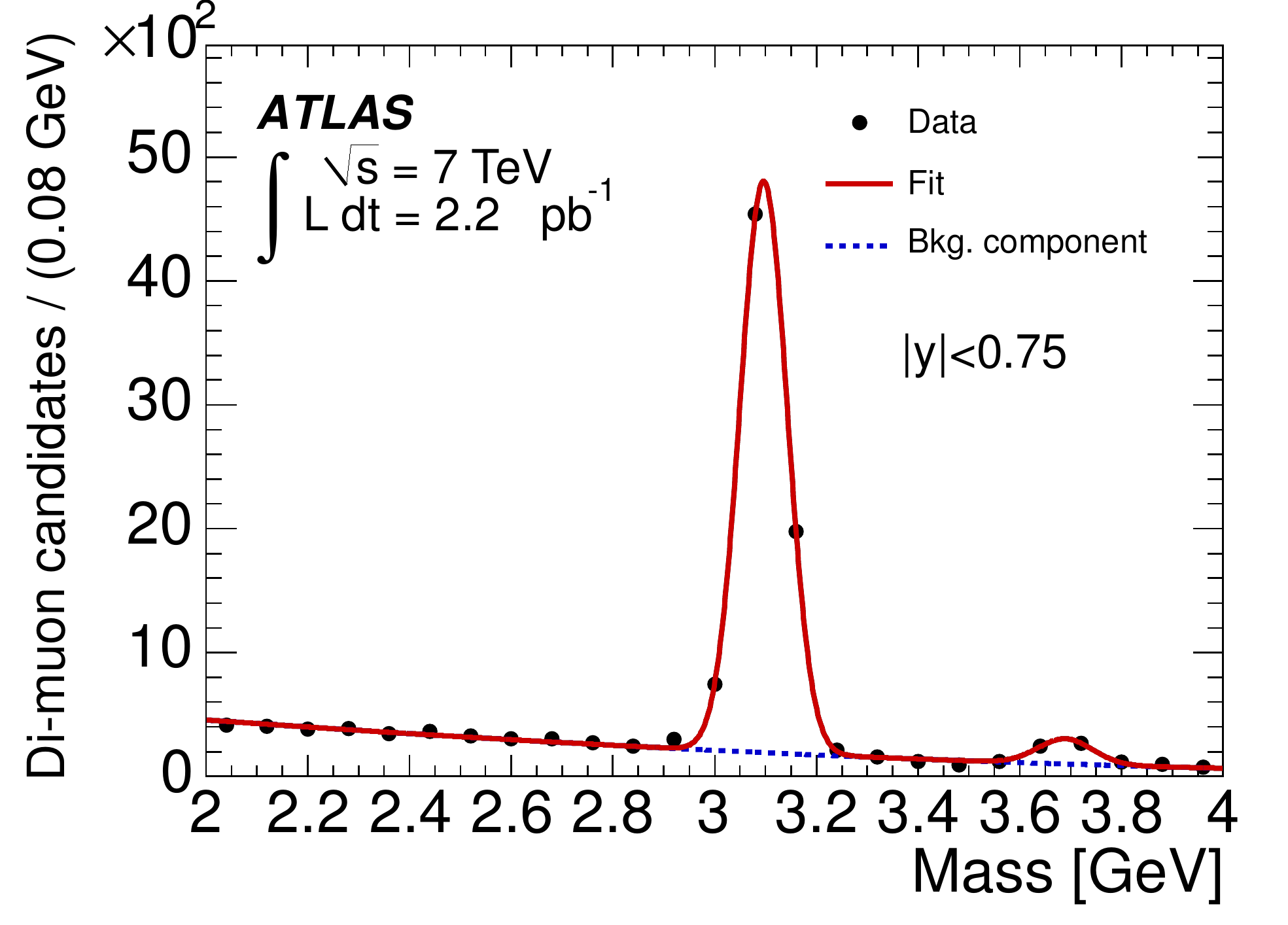}
\includegraphics[width=0.5\textwidth]{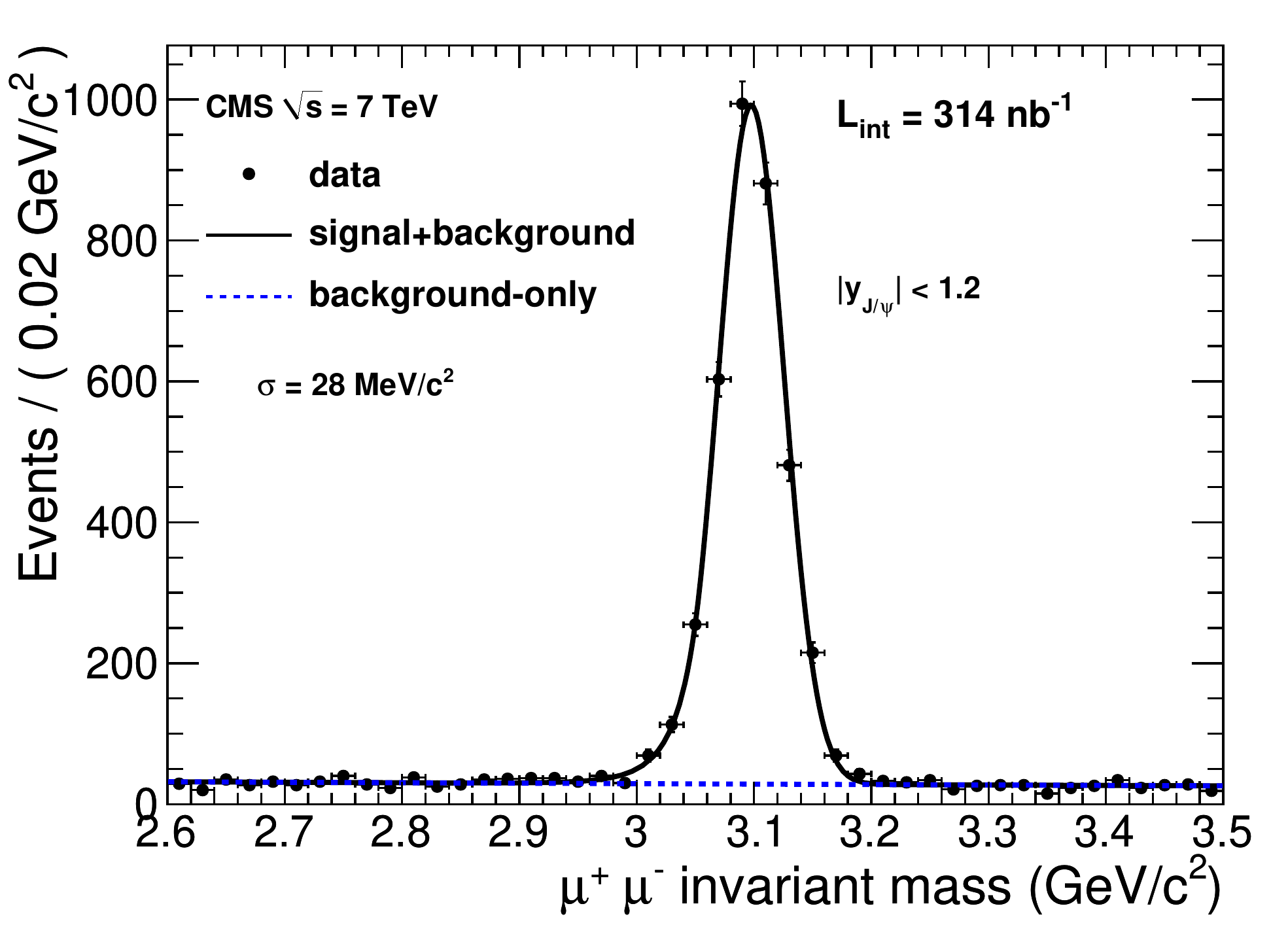}
\caption{ \jpsi\ candidates reconstructed in the barrel region of the ATLAS detector (left)~\protect\cite{Aad:2011sp} and of the 
CMS detector (right)~\protect\cite{Khachatryan:2010yr}. }
\label{fig:CMSPerformance}
\end{figure}

Both  ATLAS and CMS  are designed as multi-purpose experiments, focusing mainly 
on searches for new physics phenomena at large transverse momentum scales. 
Both experiments cover
a rapidity range up to $|\eta|< 2.5$. However, the bulk of the b-quark production peaks at large rapidities and
both experiments are only able to collect a fraction of the produced B-hadrons.
The main sub-detectors used for heavy-flavor measurements
are the tracking devices located closest to the interaction point and the muon-detectors, the sub-detector farthest out. 
The muon detectors are used to identify and to trigger on muon decays of B-hadrons.
The tracking devices consist of high precision silicon detectors, and they are used to reconstruct the
production and the decay vertices of B-hadrons. The large boost of high momentum B-hadrons leads to 
secondary decay vertices dislocated up to several millimeters from the 
primary production vertex. The tracker and the muon systems are placed in a large magnetic field to
determine the momentum of the charged tracks with high accuracy, leading
to a concomitantly high mass resolution.  Fig.~\ref{fig:CMSPerformance}
shows the invariant mass distribution of \jpsi\ candidates decaying into two muons being 
reconstructed in the barrel region of the ATLAS detector~\cite{Aad:2011sp}
and the CMS detector~\cite{Khachatryan:2010yr}. The mass resolution is determined to 
be  between $\sigma$ = 46 to 111  MeV (ATLAS) and between $\sigma $= 20 and 50  MeV (CMS),
broadening with increasing rapidity. 
The high mass resolution is important  to reach a good signal to background ratio for reconstructed 
B-hadron candidates. 
The impact parameter resolution is determined by the single hit resolution of the tracking devices and is 
limited by multiple scattering effects of low momentum tracks. For any measurement using lifetime based quantities 
the impact parameter measurement is a crucial ingredient. \par
It should be noted that both detectors have only very limited particle identification possibilities.
Only the silicon tracking devices have some kaon-pion separation power in the 
momentum region below one GeV/$c$. Above one GeV/$c$ no kaon-pion separation is 
possible, leading to increased background conditions for reconstructed B-hadron 
candidates with kaons in the final state. 
The muon detectors are essential for selecting B-hadron events.
Since tracking devices are not part of the first trigger decision, B-hadron decays
with leptons in the final state are used to select B-hadron events. The most promising possibility offer 
B-hadron decays with two oppositely charged muons in the final state like  
the decay into a \jpsi\ or B-hadron decaying directly into
a pair of muons. 
However, the overall trigger rate with two muon candidates in the final state exceeds the limits imposed by both the 
trigger itself and the long-term data storage resources, so this must be mitigated by applying trigger pre-scales or by 
raising the momentum thresholds for the candidates (with a preference for the latter).
Overall the ATLAS detector collects  5.25  fb$^{-1}$ and the CMS detector 5.56 fb$^{-1}$ of data in 2011 with a maximum
instantaneous  luminosity of $\sim3.5 \times 10^{33} \mathrm{ cm^{-2} s^{-1}}$. 
For this luminosity the setup of the accelerator induces on average up to 11.6 collisions 
per bunch crossing, a challenge for any heavy flavor physics measurement.  
\section{B-Hadron Production}
\label{BHadProd}
In a hadron collider such as the LHC the bulk of b-quarks and the following B-hadrons are produced 
via the strong interaction. The production mechanism can be described in the context of 
perturbative and non-perturbative Quantum Chromo Dynamics (QCD) and a measurement 
of the B-hadron production rate offers an excellent  test of QCD models. Besides the production 
of B-hadrons the generation of quarkonium states like the \us, \utwos, \uthrees, the \jpsi\ and the \jpsitwos\ are also
of great interest. Quarkonium production measurements can be used to probe the mechanism of $\mathrm{q\bar{q}}$-generation
followed by the subsequent evolution of the quark pair into a quarkonium state~\cite{Brambilla:2010cs}.
However, B-hadron decays and higher mass charmonium states like the \chic\ will
contribute to the overall non-prompt charmonium production rate.
\subsection{Quarkonium Production and Observation of a new \chib\ State}
\begin{figure}[ht]
\includegraphics[width=0.42\textwidth]{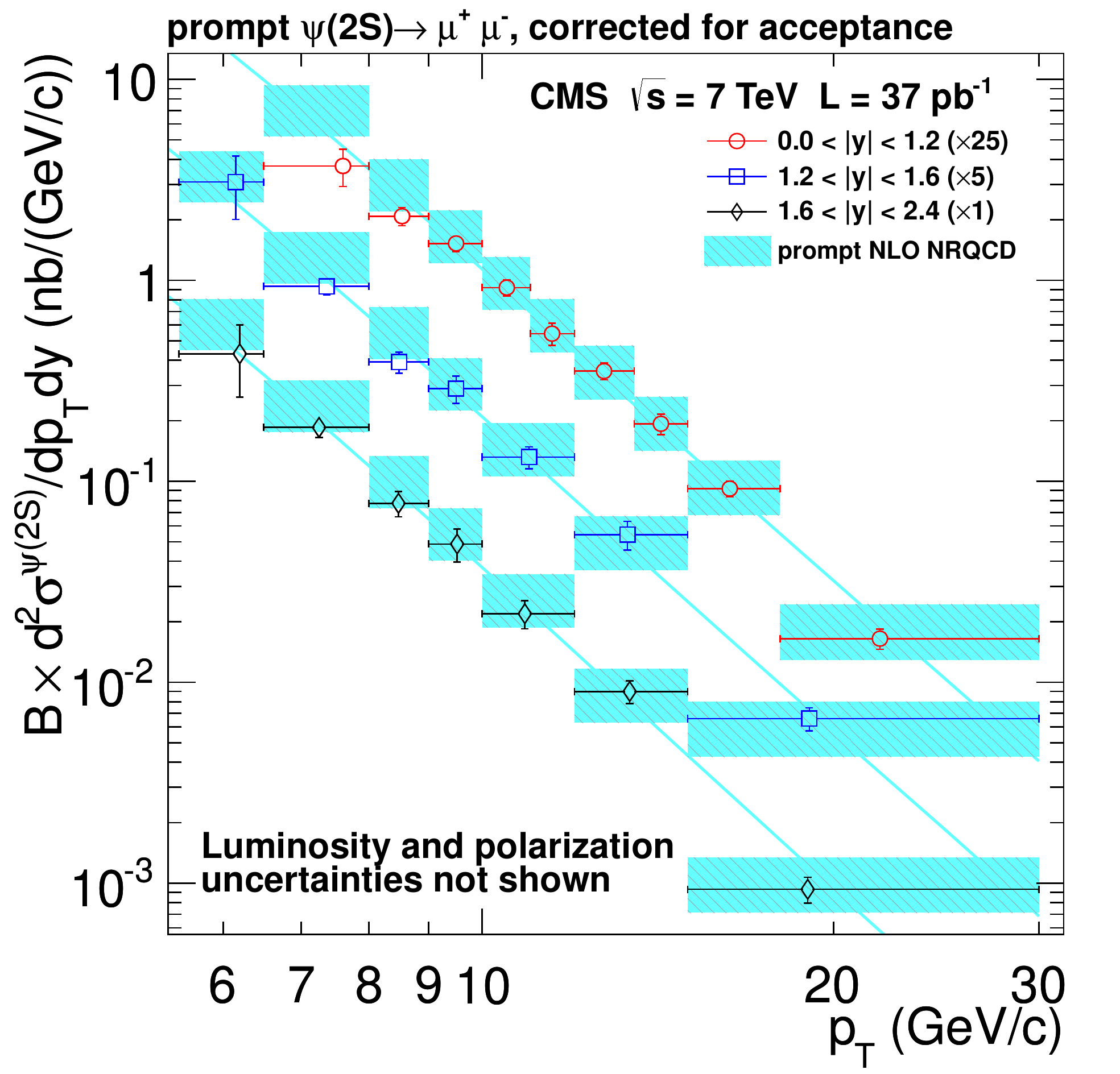}
\includegraphics[width=0.58\textwidth]{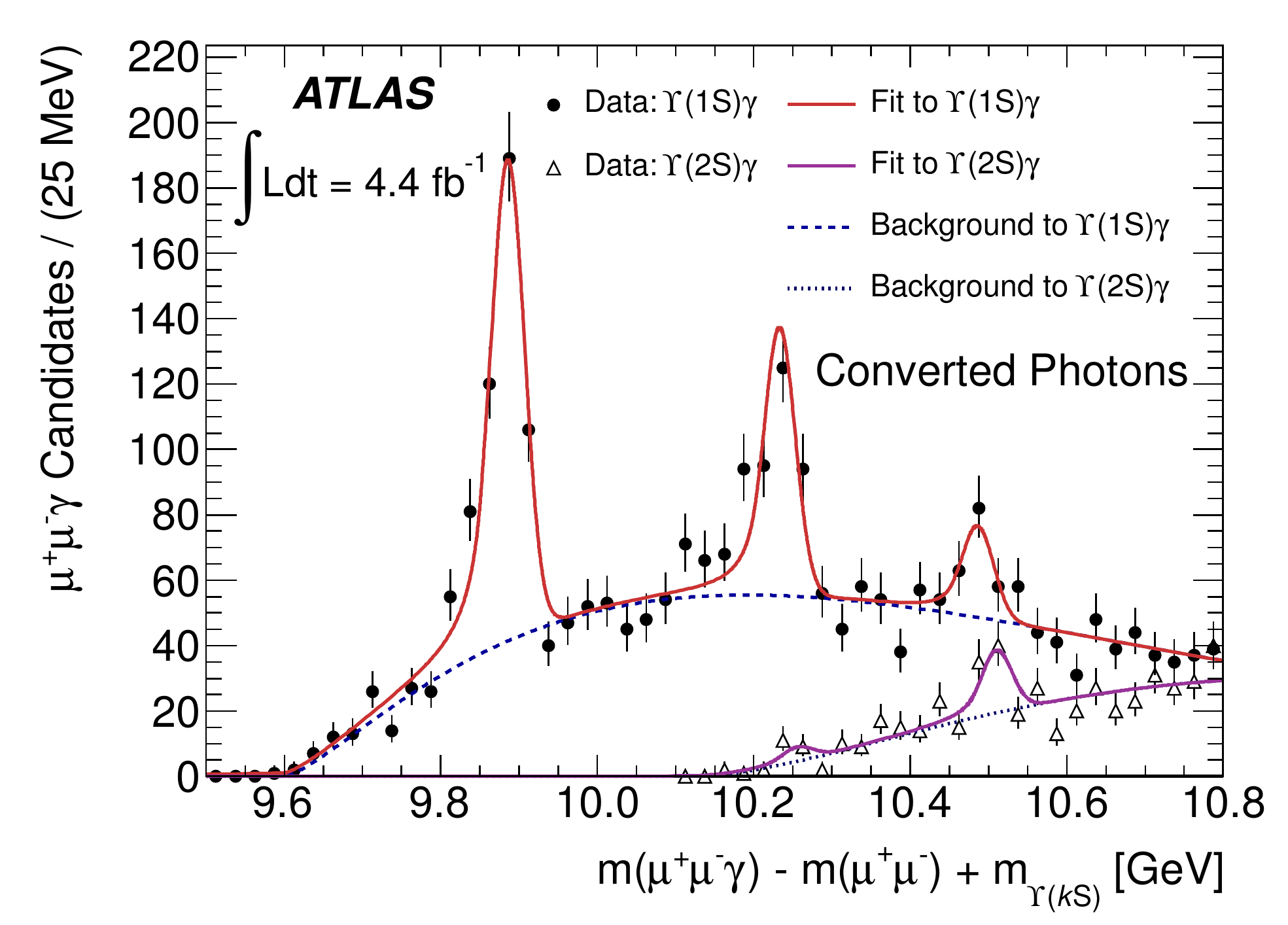}
\caption{The measured differential cross section as measured by the CMS Collaboration 
is shown in the left plot for prompt \jpsitwos\ as a function of \pt\ and for different rapidity bins~\protect\cite{Chatrchyan:2011kc}. 
The right plot shows the mass distribution of \chidecay\ (k=1,2) candidates 
reconstructed with the ATLAS detector with the photon being converted and reconstructed by two electrons~\protect\cite{Aad:2011ih}. The circles
show the data for the \chib\ decaying into a \us\  and a photon and the triangles correspond to the \chib\ decaying 
into a \utwos\ and a photon.}
 \label{fig:chib} 
\end{figure}

Precise measurements of quarkonium production are performed by the ATLAS and the CMS Collaborations.
Both collaborations measure the charmonium production rate and compare it to 
various theoretical predictions. The CMS Collaboration~\cite{Chatrchyan:2011kc} studies the 
prompt and non-prompt production rate of \jpsi\ and \jpsitwos\ as a function of \pt\ and rapidity. The
data sample corresponds to an integrated luminosity of 37 pb$^{-1}$.
The prompt (non-prompt) production rate is compared to NLO NRQCD (FONLL) theory predictions. All measurements 
are in agreement  with the theory predictions. The agreement found for the prompt \jpsitwos\ production rate 
is of particular interest since no feed-down from higher mass charmonium states is expected and therefore 
a direct  comparison to the theory prediction is possible.  The differential production rate  of the prompt \jpsitwos\  production rate 
as a function of \pt\ in several rapidity bins is shown in Fig.~\ref{fig:chib}.
The ATLAS Collaboration measures the prompt and non-prompt \jpsi\ production rate (using 2.3 pb$^{-1}$ of data)
as a function of \pt\ and compares the prompt production rate to a simple color evaporation model and  
NLO and partial NNLO color singlet models (CSM)~\cite{Aad:2011sp}. The partial NNLO CSM
shows a significant improvement in describing the \pt\ dependence and the normalization
of the prompt production rate, however, differences are still visible.
The ATLAS Collaboration performs a measurement of the \us\ production cross 
section based on 1.13 pb$^{-1}$ of data~\cite{Aad:2011xv}.  The \us\ production rate is  dominated by prompt production
and contributions from higher mass decays are negligible. The differential production rate as a function
of \pt\ is compared to a NLO CSM with direct contributions only, and to a NRQCD model
implemented in PYTHIA~8. The data exceeds significantly the NLO CSM, similar to the charmonium case, which could be explained
by the need for higher order corrections. The agreement with the NRQCD predictions are reasonable, but differences
up to a factor of two are observed. The CMS Collaboration measures with 3.1 pb$^{-1}$ of integrated luminosity 
the total and differential production cross section
of the \us, \utwos\ and the \uthrees~\cite{Khachatryan:2010zg} and comparisons to predictions from PYTHIA are performed. 
The normalized differential cross section with respect to \pt\ obtained with PYTHIA is consistent with the measurements, while the
overall cross section is overestimated. \par
The ATLAS Collaboration reports on the observation of a new \chib\ state in its radiative transition to
 \us\ or  \utwos~\cite{Aad:2011ih} using an integrated luminosity of 4.4 fb$^{-1}$. The bottomonium states \chib(1P) and \chib(2P) are previously observed 
experimentally  and the existence of the \chib(3P) state is expected. 
ATLAS reconstructs the \chib(3P) radiative decays from the photon emitted 
during the transition, and the subsequent decay of the \us\ or \utwos\ into two muons.
The photon is either reconstructed directly in 
the electromagnetic calorimeter or via two electrons originating from the conversion of the photon. 
The reconstructed mass of m$_{\mathrm{\chib(3P)}} = 10.530 \pm 0.005 \mathrm{\, (stat.)} \pm 0.009 \mathrm{\, (syst.)}$ GeV
is consistent with the expectation from theoretical models averaging the mass over the three \chib(3P) hyperfine triplet states. 
The mass distribution of the \chidecay\ (k=1,2) candidates with the photon reconstructed via  photon conversion
is shown in Fig.~\ref{fig:chib}.
\subsection{B-Meson and B-Baryon Production}
\begin{figure}[ht]
\includegraphics[width=0.5\textwidth]{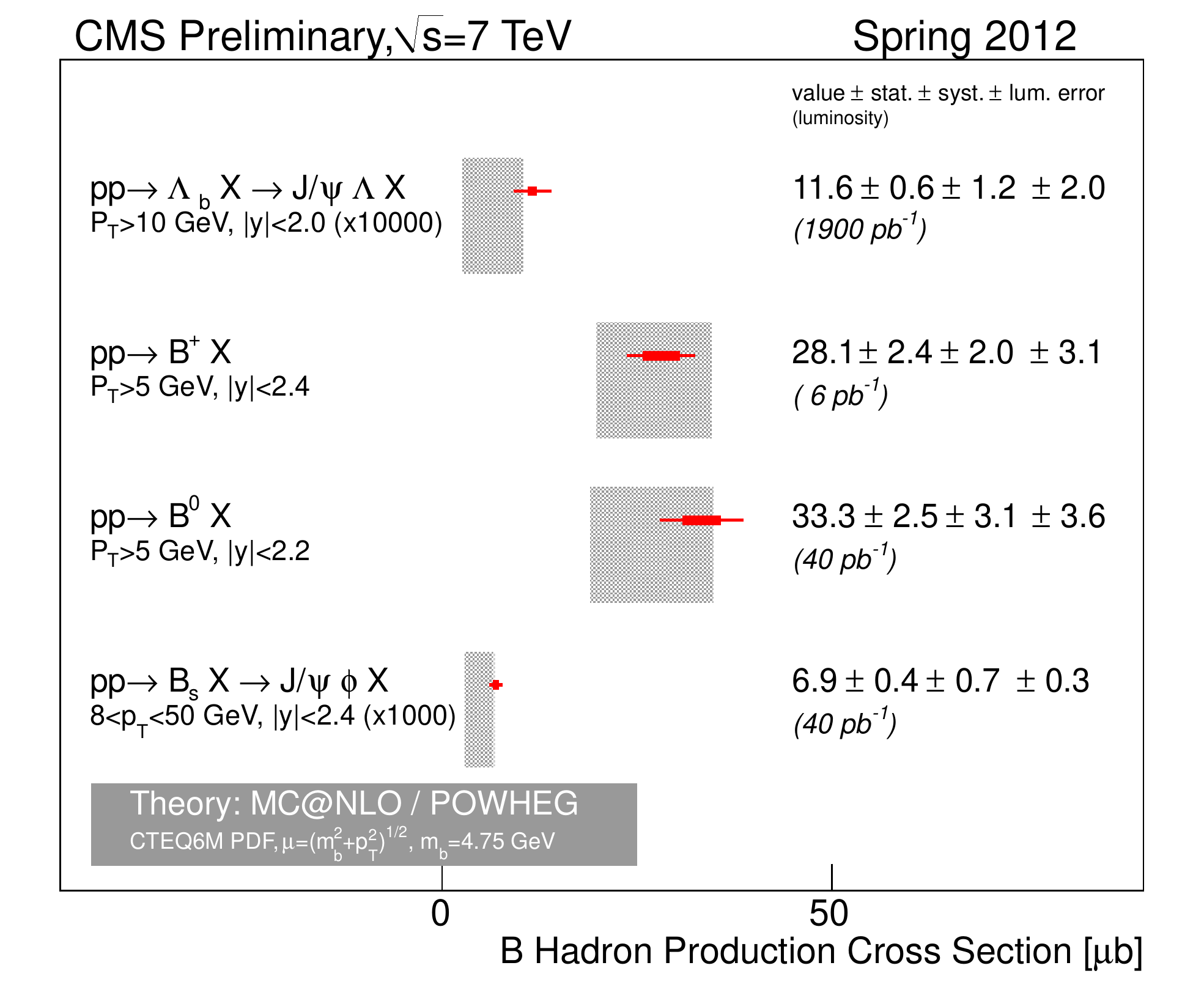}
\includegraphics[width=0.5\textwidth]{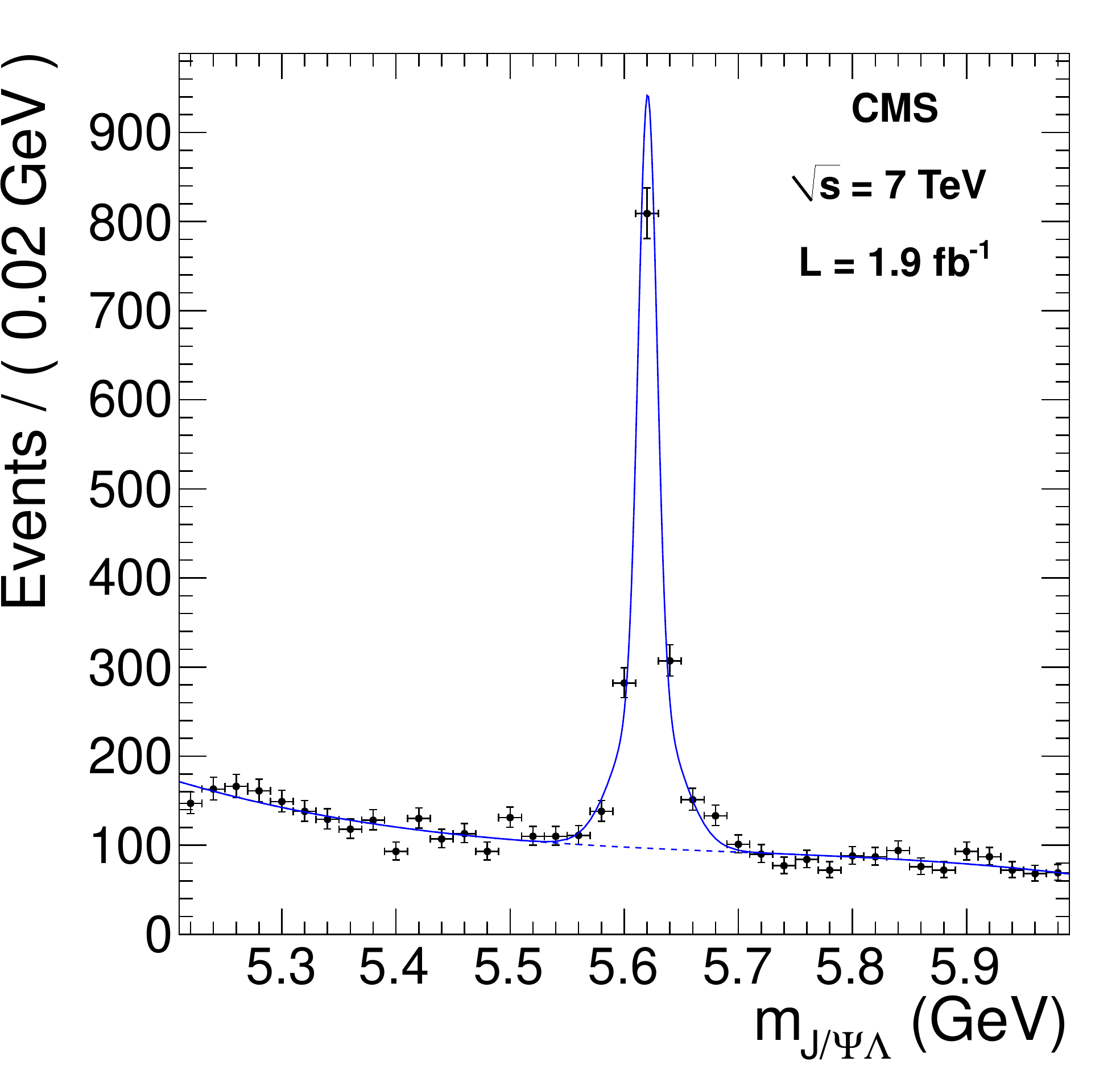}
\caption{
The B-meson cross section measurements performed by the CMS Collaboration are summarized and compared to theory
predictions (MC@NLO) in the left figure. The inner error bar corresponds to the
statistical uncertainty, the outer error bar to the quadratic sum of statistical and systematic uncertainty. 
The outermost bracket is the total uncertainty, including
the luminosity uncertainty. 
The invariant mass distribution of the \lambb\ reconstructed  in the decay \lambdab\ is shown on the right~\protect\cite{CMS:lambdaB}. 
The selected   \lambb\ candidates are reconstructed with a \pt\ larger than 10 GeV/$c$ and a rapidity smaller than two.
}
\label{BMesonSummary}
\end{figure}

\begin{figure}[ht]
\includegraphics[width=0.5\textwidth]{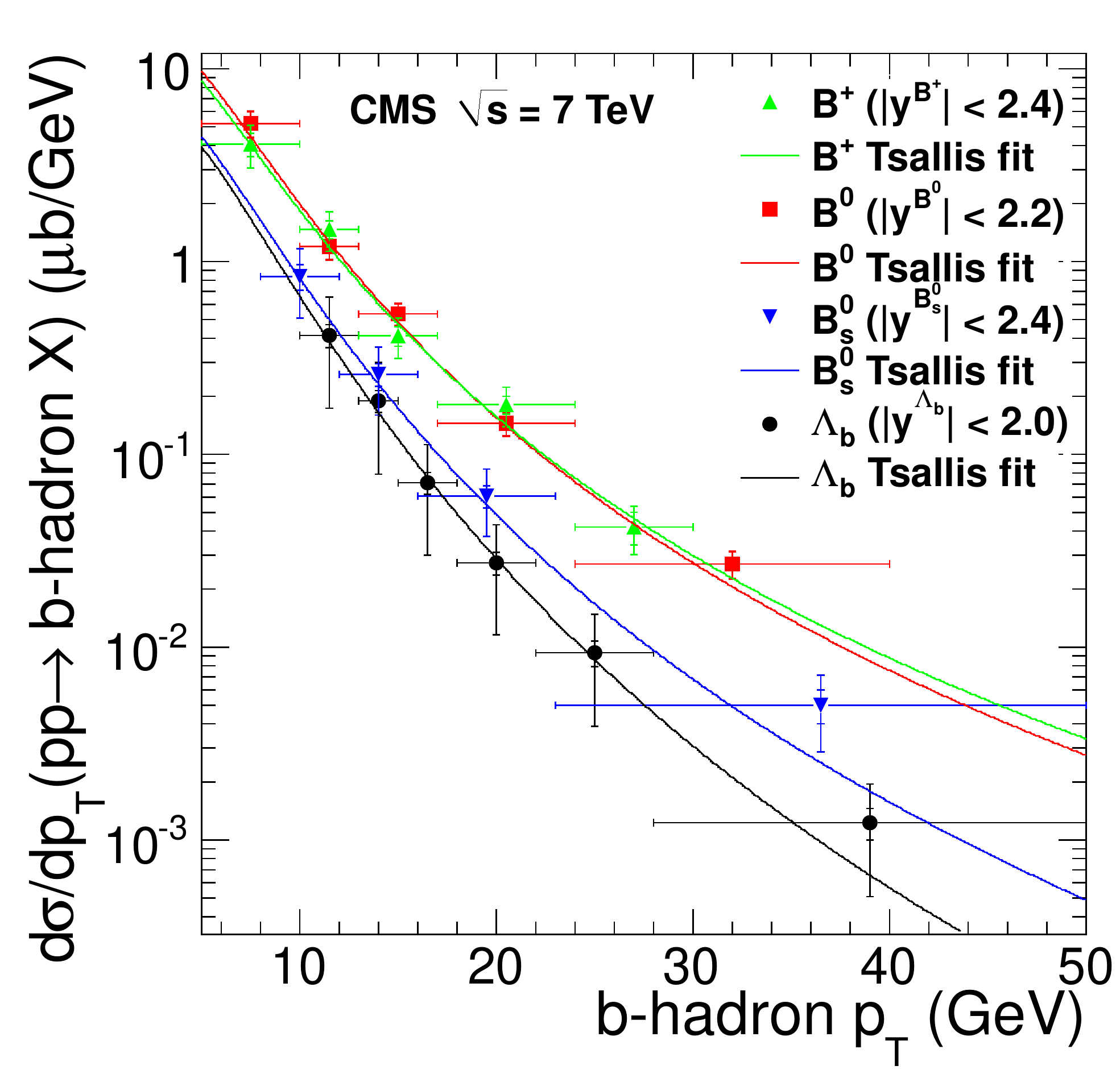}
\includegraphics[width=0.5\textwidth]{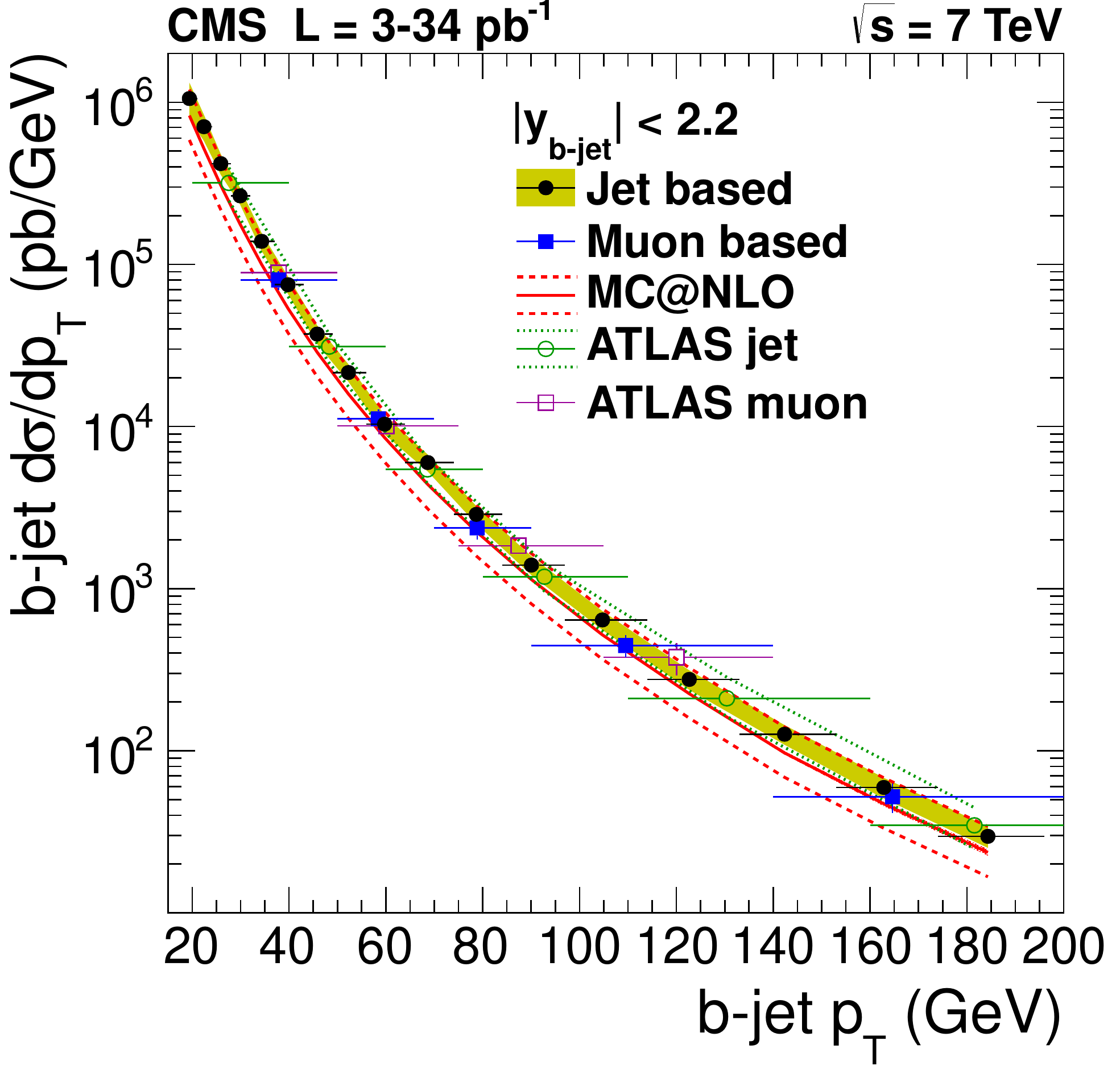}
\caption{The distribution on the left shows the differential cross section versus \pt\ reconstructed 
for the exclusive decays
of the  \bpm-meson, \bd-meson and \bs-meson and the \lambb-baryon~\protect\cite{CMS:lambdaB}.
On the right the inclusive b-jet production rate measured by the ATLAS and CMS Collaboration 
together with a comparison to MC@NLO is shown~\protect\cite{Chatrchyan:2012dk}.}
\label{fig:lambdabpt} 
\end{figure}

The CMS Collaboration measures the total cross section 
of the \bpm-meson, \bd-meson and \bs-meson using 
exclusively reconstructed decay channels  $\bpm\to\jpsi \mathrm{K}^\pm$ (using 5.8 pb$^{-1}$),  
\bjpsiks\ and \bsjpsiphi\ (both using 40 pb$^{-1}$)~\cite{Chatrchyan:2011vh,Chatrchyan:2011pw,Khachatryan:2011mk}. Besides the total cross section the 
differential production cross section is determined as a function of the transverse momentum \pt\ and the rapidity of the B-meson. The measured cross sections are compared to MC@NLO and PYTHIA Monte Carlo predictions. 
For all three cross section measurements the theoretical prediction by MC@NLO is below the measured 
value while the PYTHIA prediction is above the measured one. The rapidity dependence observed in data is flatter than predicted 
by PYTHIA. The \pt\ spectrum falls slightly faster than predicted by MC@NLO. A summary of the total cross section measurements  
and a comparison to MC@NLO is summarized in Fig.~\ref{BMesonSummary}. 
CMS also releases a new measurement  of the \lambb\ production cross section in pp-collisions at 7 TeV
using an integrated luminosity of 1.9 fb$^{-1}$~\cite{CMS:lambdaB}. 
The invariant mass distribution of the \lambb\ exclusively
reconstructed with the decay \lambdab\ is shown in Fig.~\ref{BMesonSummary}. The cross section measurement is compared to a
 prediction from PYTHIA. The production rate as a function of the rapidity as well as the ratio $\mathrm{\bar \lambb /\lambb}$ is well reproduced by 
PYTHIA. The \pt\ spectrum of the \lambb\ falls faster than the  observed  meson production, 
triggering the question as to whether there is a \pt\ dependent hadronization effect in the ratio of baryon to meson production. A comparison 
of the \pt\ spectrum of the b-mesons and the \lambb-baryon is summarized in Fig.~\ref{fig:lambdabpt}.
The CMS Collaboration also presents a measurement of the inclusive b-jet production rate at the LHC~\cite{Chatrchyan:2012dk}. 
The results are consistent with the production rate measured by ATLAS and the Monte Carlo predictions using MC@NLO. 
The inclusive b-jet production rate is shown in Fig.~\ref{fig:lambdabpt}.
\section{Lifetime Measurements}
\label{Lifetime}
The excellent performance of the tracking devices allows precise measurements of the lifetime of B-hadrons. 
Besides the measurement of the average B-hadron lifetime a time dependent angular analysis of the 
decay \bsjpsiphi\ allows a measurement of the \bs-lifetime difference \dgs\ of the heavy and light mass eigenstate 
of the \bs\ meson. 
The ATLAS Collaboration presents a measurement of the average B-hadron lifetime using 
the data taken in 2010, which corresponds to about 40 pb$^{-1}$~\cite{BAverageLifetime}. The average B-hadron lifetime
is determined via the B-hadron decaying into a $\jpsi$+X, with the $\jpsi$ subsequently 
decaying in a pair of muons. The decay length of the $\jpsi$ is reconstructed and a correction factor is applied to
deduce the lifetime of the parent B-hadron. The correction factor is determined from 
simulated events, weighted to match the measured \jpsi\ momentum distribution. The  average
B-hadron lifetime is measured to be $<\tau_{b}> = 1.489\pm 0.016 \mathrm{(stat.)} \pm 0.043 \mathrm{(syst.)}$ ps.
The systematic uncertainty is dominated by the lifetime model of the background events and the
alignment uncertainty of the tracking detector. \par
\begin{figure}
\includegraphics[width=.5\textwidth]{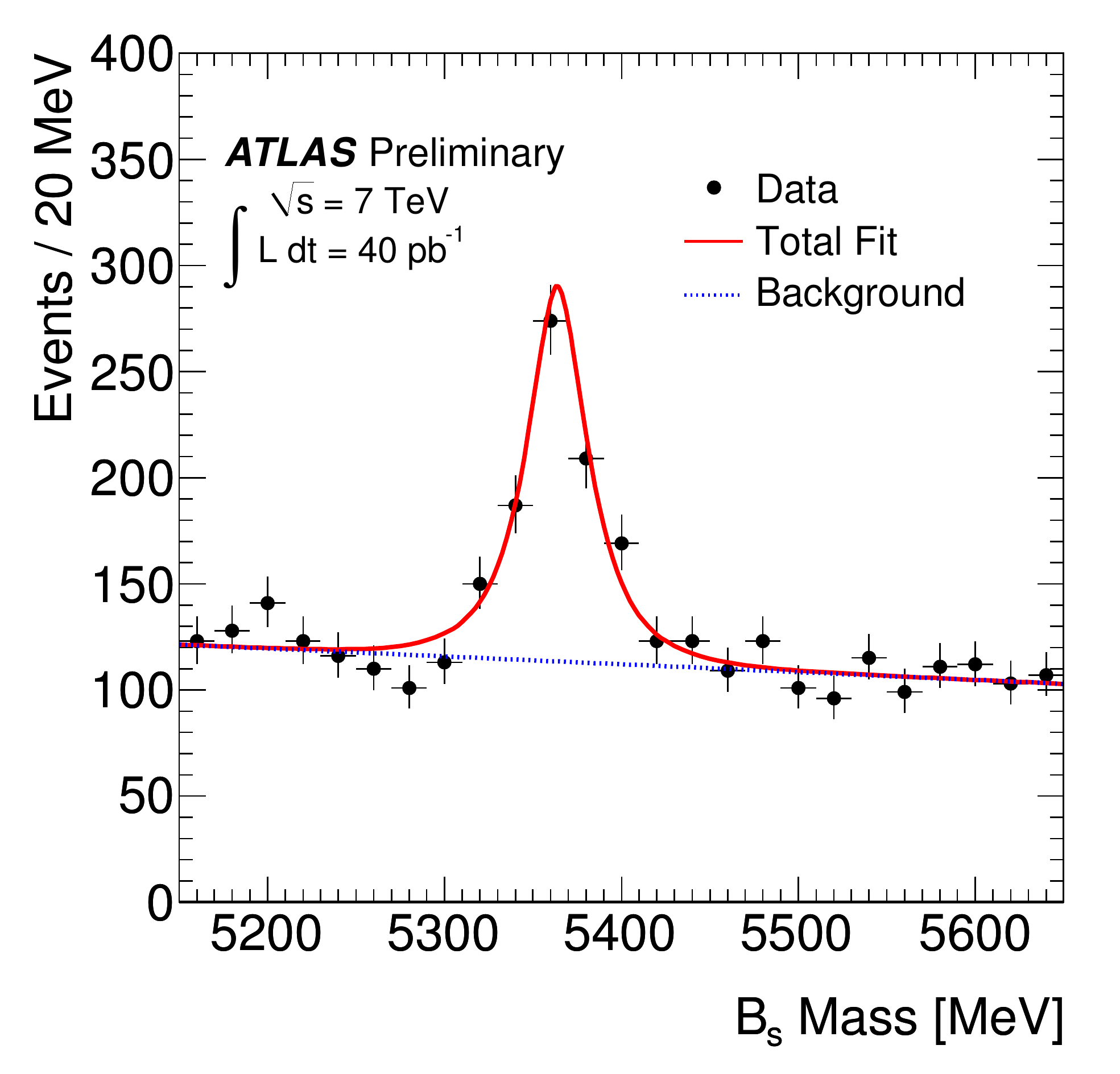}
\includegraphics[width=.5\textwidth]{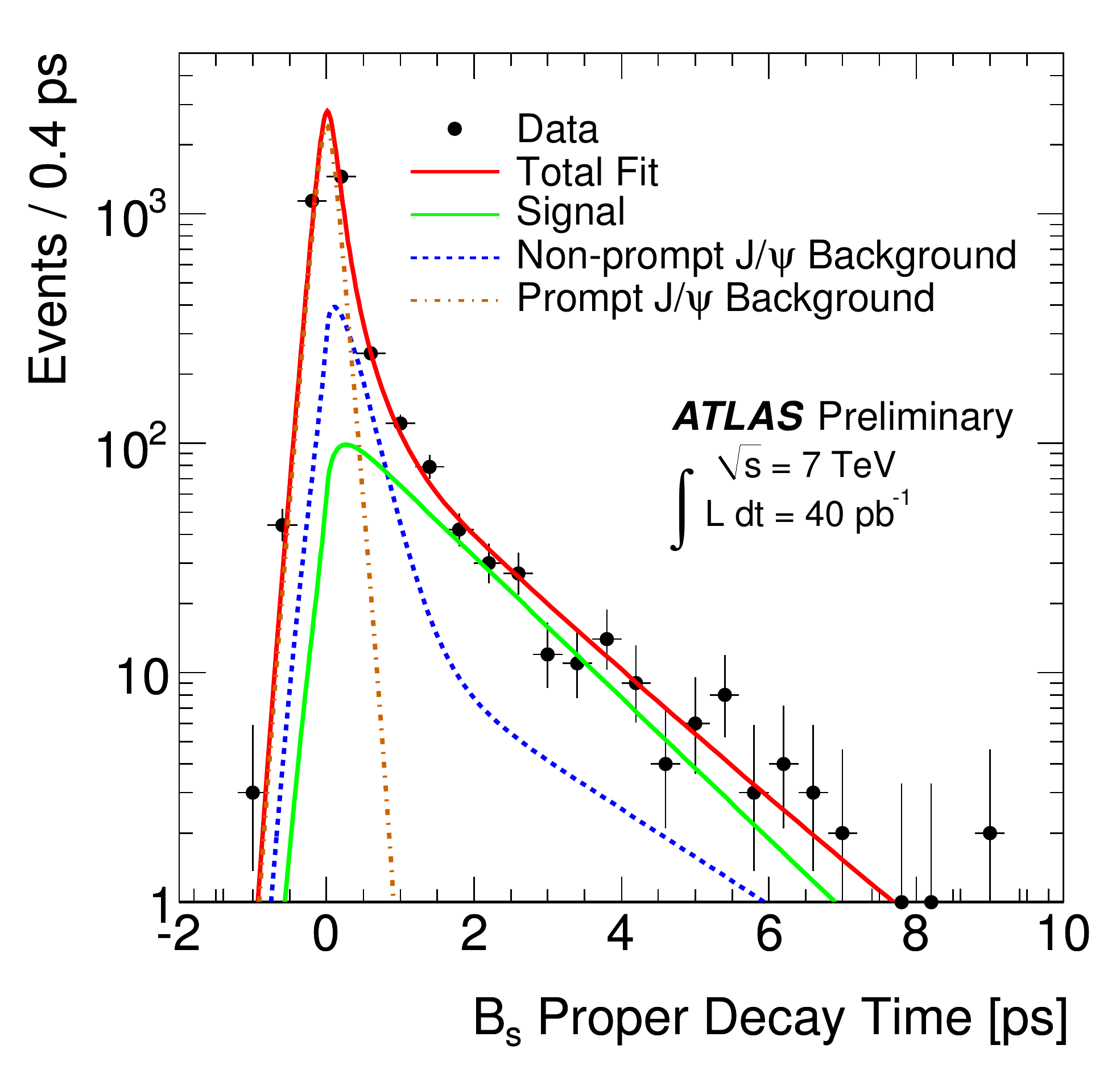}
\caption{The left plot shows the mass distribution of the reconstructed \bsjpsiphi\ events~\protect\cite{BsJPsiLifetime}. The 
right plot shows the lifetime distribution~\protect\cite{BsJPsiLifetime}. The solid lines are the projections of a simultaneous
fit to the mass and lifetime distribution, with the red line reflecting the total fit, the dotted blue
line the background contribution and the green line corresponds to the lifetime distribution of
signal events. The red dash-dotted line is the contribution from prompt \jpsi\ events. 
}
\label{fig:JPsiPhiLifetime}
\end{figure}

 
A measurement of the average lifetime of the \bs\ meson using exclusively reconstructed
\bsjpsiphi\ decays is carried out as well, using the same integrated luminosity of 40 pb$^{-1}$~\cite{BsJPsiLifetime}. 
Overall $463\pm26$ signal events are reconstructed 
with estimated $714\pm38$ background events. The single lifetime
is measured to be $\tau_{B_{s,single}}=1.41\pm0.08\mathrm{(stat.)}\pm 0.05 \mathrm{(syst.)}$ ps.
Similar to the average lifetime measurement the systematic uncertainty is dominated by the
lifetime model of the background as well as the alignment uncertainty of the tracker. Fig.~\ref{fig:JPsiPhiLifetime}
shows the mass and the lifetime distribution of the reconstructed \bsjpsiphi\ events.
In both cases the precision of the lifetime measurement is limited by the incomplete knowledge of the tracking
detector alignment. It is expected that with increasing statistics and improved knowledge of the tracking devices
the uncertainty will decrease significant. The single lifetime measurement of the 
using \bsjpsiphi\ events is an important step towards the measurement of the weak phase
$\phi_{s}$ and the lifetime difference \dgs.
\section{Rare Decays}
\label{RareDecays}
\begin{figure}[ht]
\includegraphics[width=.5\textwidth]{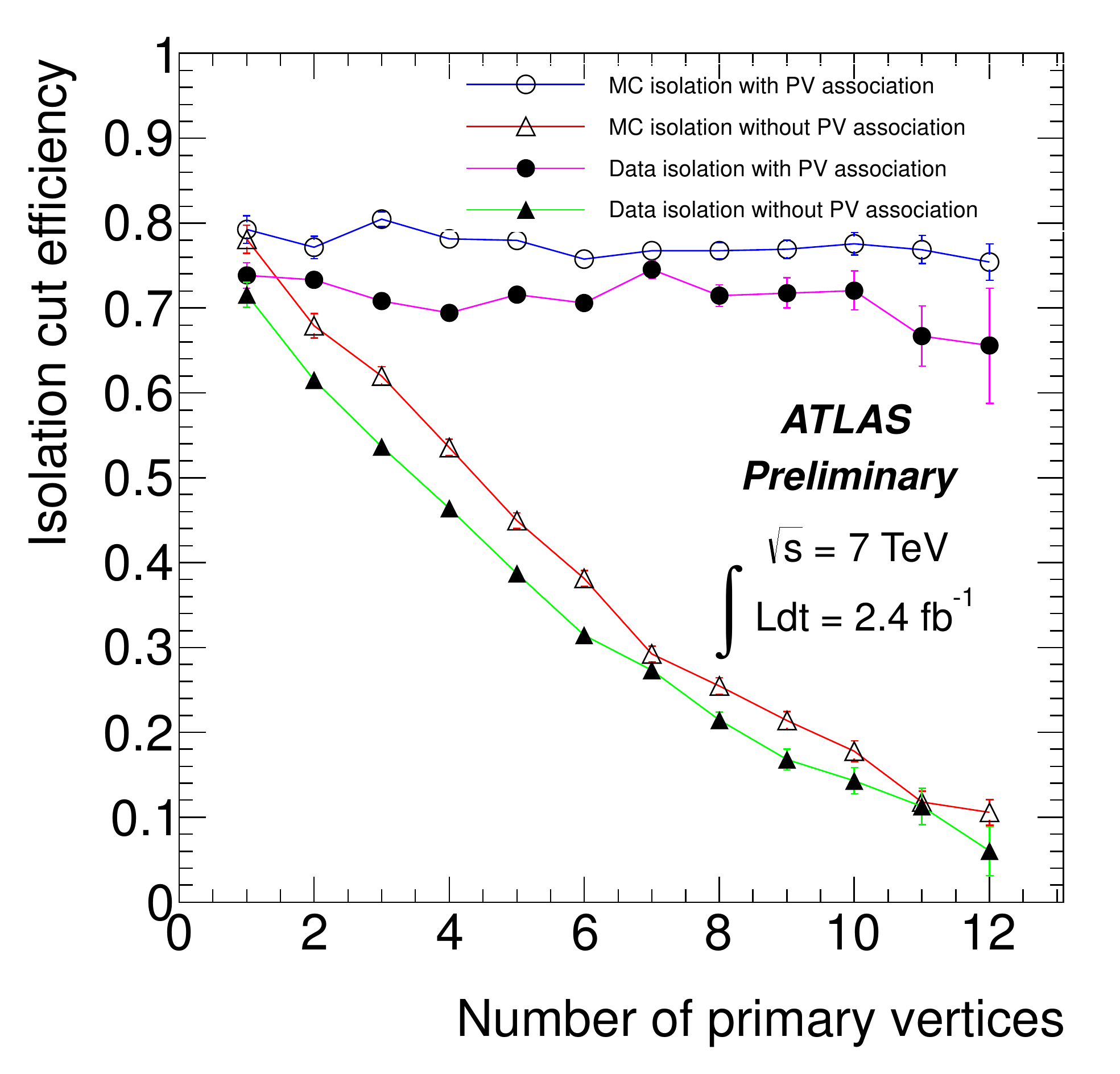}
\includegraphics[width=.5\textwidth]{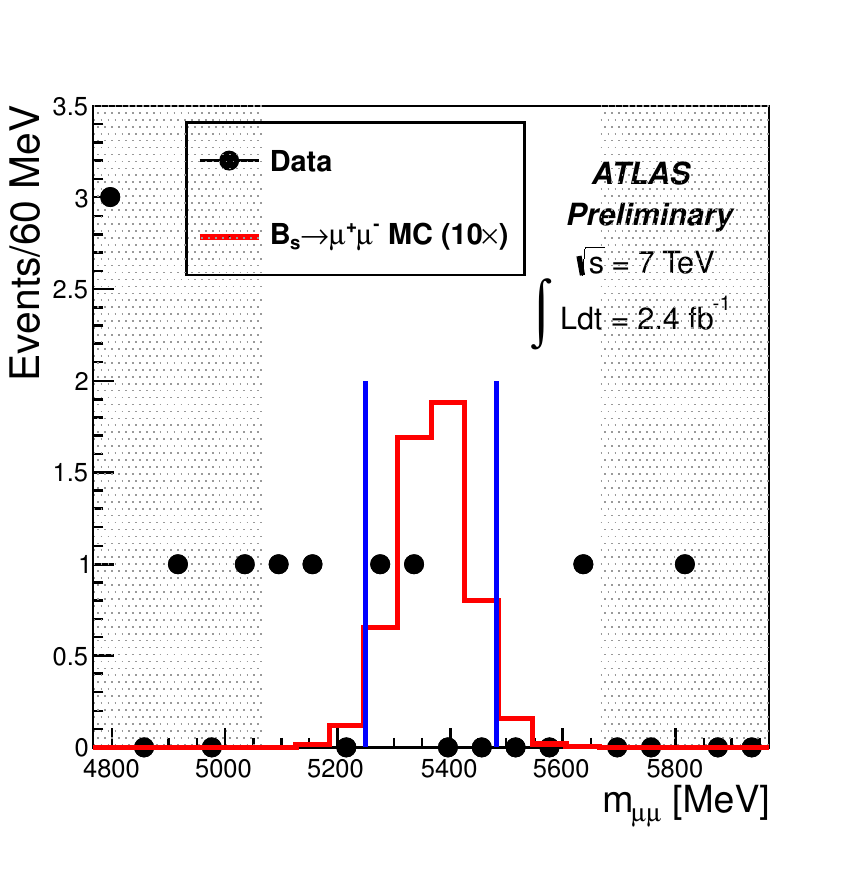}
\caption{The left plot shows the isolation cut efficiency as a function of reconstructed primary vertices~\protect\cite{RareBATLAS}.  The triangles
show the efficiency for data (closed) and simulated events (closed)  not associating the events to the primary vertex
as a function of reconstructed primary vertices. Using the primary vertex information the isolation cut efficiency
is independent of the number of reconstructed primary vertices,  for data (close circles) as well as for simulated events (open 
circles). The right plot shows the number of reconstructed candidates as a function of the invariant mass, where both muons
are reconstructed in the barrel region of the detector~\protect\cite{RareBATLAS}. The blue line indicate the signal region, the grey area corresponds to the sideband
and the red histogram indicates the expected number of signal events multiplied by a factor of ten. }
\label{fig:Bsmumu}
\end{figure}

The study of rare decays like \bdmumu\ and \bsmumu\ offers the possibility fo searching for physics beyond the Standard Model. 
This approach is an alternative way to the direct searches performed at ATLAS and CMS selecting 
high \pt\ events. Theoretical calculations predict a branching ratio of $3.5\pm0.3 \times 10^{-9}$ 
for the decay \bsmumu.
The ATLAS, CMS and LHCb Collaborations present new measurements on rare 
B-hadron decays~\cite{RareBATLAS,Chatrchyan:2012rg,Aaij:2012ac}. The analysis
performed by CMS and LHCb use the complete data set collected in 2011 and are 
discussed in a separate presentation. In this paper the focus will be on the ATLAS measurement 
searching for the rare decay \bsmumu\ using data taken with the ATLAS experiment in 2011.
Since the trigger conditions changed  during the data taking in 2011 the analysis is performed
with 2.4 fb$^{-1}$ of collected data only. 
The branching ratio is not determined directly, but by measuring the ratio to a more abundant B-hadron decay 
$\bpm \to \jpsi$K$^{\pm}$. 
The efficiency, acceptance and the hadronization probability for a b-quark transforming to a B-hadron 
is different for both decays. The differences are estimated using simulated and data events and
the branching ratio \bsmumu\ is finally determined by using the measured ratio.
Signal events are separated from
background events using a multivariate classifier with 14 discriminating variables. During 
data taking the event characteristics changed significantly, in particular the number of reconstructed primary vertices per bunch 
crossing. The left plot of Fig.~\ref{fig:Bsmumu} shows the efficiency dependence of one of the discriminating
variables, the isolation cut which described the activity around a selected $B$-hadron candidate, as a function
of reconstructed primary vertices.  After assigning the tracks to the primary vertex the efficiency shows
no dependency on the number or reconstructed primary vertices. This clearly shows that this search for
rare decays can be still performed with increased instantaneous luminosity.
Four out of these 14 discriminating variables are used to optimise the multivariate selection.
Two main sources of background events are expected - a continuum background, which is expected to have a smooth dependence 
on the invariant di-muon mass and a resonant background of mis-reconstructed events, peaking in the same
mass range as the signal events. The continuum background can be estimated from the mass sidebands, while 
the background originating from mis-reconstructed events is estimated using simulated events. The overall 
background is dominated by the continuum background. The mass resolution varies between candidates
with both muons reconstructed in the barrel region of the detector and candidates with at least one muon 
reconstructed in the forward region of the detector. In order to take this effect into account, the search
is divided into three candidate categories depending on the direction of the most forward muon. The 
selection optimization and the background estimate from sideband events is performed with different
event samples. This choices guarantees a bias free estimate of the expected events in the signal region. 
In total 6.4 events are expected while three events are selected after the search region is unblinded. The right plot in Fig.~\ref{fig:Bsmumu}
shows the number of observed and estimated candidates in the sideband and in the signal region. 
The number of estimated background events and
the number of observed signal events together with the event reconstruction efficiency, the acceptance and the hadronization efficiency
is used to calculate the upper limit of the branching ratio. 
The expected
$95\%$ confidence limit is $2.3^{1.0}_{-0.5}\times10^{-8}$ and the observed one is $2.2\times10^{-8}$.
\section{Summary and Conclusion}
\label{summary}
We report on heavy flavor measurements using data taken in 2010 and 2011  by the ATLAS and the CMS experiment.
The large available data set of more than 5 fb$^{-1}$ per experiment and the excellent detector performance allows each 
to perform many competitive measurements. The measurements of heavy quark production cross sections allow
precision studies of QCD. Studies with heavy quarkonium states like the charmonium
states \jpsi\ and  \jpsitwos\ and the bottomonium state \us, \utwos\ and \uthrees\  are presented. The measurements are compared to
theory models and in several cases large deviations are found. A new bottomonium state, the \chib(3P), is observed for the 
first time by the ATLAS experiment. The CMS Collaboration studies the production of b-flavored hadrons and differences
to theory predictions for the total as well as differential production cross section are found. The ATLAS Collaboration presents 
an average B-hadron lifetime as well as a lifetime measurement using exclusive \bsjpsiphi\ decays. These measurements
reflect the good detector performance and are important towards lifetime dependent studies, like the measurement
of the weak mixing phase $\phi_{s}$. For the first time ATLAS presents a search for the rare decay \bsmumu. The number
of reconstructed events is consistent with background events only and a  $95\%$ confidence limit for the branching ratio
is set to $2.2\times10^{-8}$.
\section*{Acknowledgments}
This research was supported by the DFG cluster of excellence 'Origin
and Structure of the Universe'.

\end{document}